\documentclass[showpacs,twocolumn,aps]{revtex4}
\usepackage{amssymb}
\usepackage{amsmath}
\usepackage{graphicx}
\usepackage{lscape}
\usepackage{booktabs}
\RequirePackage[colorlinks,citecolor=blue,urlcolor=blue,linkcolor=blue]{hyperref}

\begin{document}

\title{Study of nuclear modification factors of (anti)hadrons and light (anti)nuclei
in Pb-Pb collisions at $\sqrt{s_{\rm{NN}}}$ = 2.76 TeV %\thanksref{t1}
}
%\subtitle{Do you have a subtitle?\\ If so, write it here}
\author{Zhi-Lei She$^{1,2}$, Gang Chen$^{2}$\footnote{Corresponding Author: chengang1@cug.edu.cn}, Feng-Xian Liu$^{1,2}$, Liang Zheng$^{2}$ and Yi-Long Xie$^{2}$}

\address{
${^1}$Institute of Geophysics and Geomatics, China University of Geosciences, Wuhan 430074, China\\
${^2}$School of Mathematics and Physics, China University of Geoscience, Wuhan 430074, China}

\date{Received: date / Accepted: date}
% The correct dates will be entered by the editor

\begin{abstract}
The nuclear modification factors ($R_{AA}$) of $\pi^{\pm}, p(\bar p)$, and $d(\bar d)$ with $|y|<0.5, p_T<20.0$~GeV/c
in peripheral (40-60\%) and central (0-5\%) Pb-Pb collisions at $\sqrt {s_{NN}}=2.76$ TeV have been studied using
the parton and hadron cascade ({\footnotesize PACIAE}) model plus the dynamically constrained phase space coalescence ({\footnotesize DCPC}) model.
It is found that the $R_{AA}$ of light (anti)nuclei ($d, \bar d$) is similar to that of hadrons ($\pi^\pm, p, \bar p$), and the $R_{AA}$ of
antiparticles is the same as that of particles. The suppression of $R_{AA}$ at high-$p_T$ strongly depends on event centrality and mass of
the particles, i.e., the central collision is more suppressed than the peripheral collision.
Besides, the yield ratios and double ratios for different particle species in $pp$ and Pb-Pb collisions are discussed, respectively.
It is observed that the yield ratios and double ratios of $d$ to $p$ and $p$ to $\pi$ are similar to those of
their anti-particles in three different collision systems, suggesting that the suppressions
of matter ($\pi^{+}, p, d$) and the corresponding antimatter ($\pi^{-},\bar{p},\bar{d}$) are around the same level.

\end{abstract}

\maketitle

\section{Introduction}
\label{intro}
It is known that quark-gluon plasma({\footnotesize QGP}), a new form of nuclear matter characterized by the deconfined state of quarks and gluons,
can be produced in heavy-ion collisions at ultra-relativistic energies,
such as at the Relativistic Heavy-Ion Collider ({\footnotesize RHIC}) at {\footnotesize BNL}
and Large Hadron Collider ({\footnotesize LHC}) at {\footnotesize CERN}. Since a large amount of energy is deposited in
the extended {\footnotesize QGP} matter, it is allowed to create abundant anti-matter ranging from hadrons to light nuclei. Quantitative studies
on the production of anti-matter in high energy heavy ion collisions will shed light on the understanding to the anti-matter to matter asymmetry in our universe.
Up to now, numerous experimental results of (anti)hadrons ($\pi^{-},\bar{p}$, $\overline{\Lambda}$, etc.) and
(anti)nuclei ($\overline d$, $\overline{^3{He}}$, and $\overline{_{\overline\Lambda}^3 H}$, etc.)
in $pp$~\cite{PLB2014196,PhysRevC.93.034913,PRC.97024615} and Pb-Pb~\cite{PLB2014196,PhysRevC.93.034913,PRC88044910,PRL111222301,PRC93024917,EPJC2017658,PLB2016360} collisions
at $\sqrt{s_{\rm{NN}}}$ = 2.76 TeV have been reported.

Transverse momentum spectra of various particle species in nucleus-nucleus (A-A) collisions can be applied to study many important
properties of the {\footnotesize QGP} matter. The microscopic process at low $p_T$ is dominant by the bulk production. In the
intermediate $p_T$ region, the baryon-to-meson ratio shows an enhancement~\cite{PRL88022301,PRL89202301,PhysRevLett.91.172302}, which is
the so called "baryon anomaly" not fully understood so far. For the inclusive particle spectra at high $p_T$, transport properties
of the {\footnotesize QGP} matter can be obtained through jet quenching~\cite{GYULASSY1990432,npb1994583,PhysRevLett.89.162301}. Experimentally,
the nuclear modification factor $R_{AA}$ is usually performed to study the jet quenching effect~\cite{PLB2014196,ADCOX2005184, ADAMS2005102,201130,201352,Chatrchyan2012}.

The $R_{AA}$, which compares the $p_{T}$ distributions of the charged particles
in nucleus-nucleus (A-A) collisions to $pp$ collisions, is typically expressed as~\cite{PhysRevC.93.034913}:
\begin{equation}
R_{AA}(p_T)=  \frac{d^{2}N^{AA}_{id}/ d\eta dp_{T}}
{\langle T_{AA}\rangle  d^{2}\sigma ^{pp}_{id} /d\eta dp_{T}}.
\label{funct0}
\end{equation}
where $N^{AA}_{id}$ and $\sigma ^{pp}_{id}$ denote the charged particles yield per event in A-A collision
and the cross section in $pp$ collision, respectively. The nuclear overlap
function $T_{AA}$ is computed based on the Glauber model~\cite{PhysRevC.77.014906}.

The study of the $R_{AA}$ plays an important role
in understanding the detailed mechanism by which hard partons lose energy traversing the medium~\cite{APPB2007}.
Recent experimental data of $R_{AA}$ in Pb-Pb collision from {\footnotesize ALICE}~\cite{PLB2014196,PhysRevC.93.034913,201130,201352,Acharya2018}
and {\footnotesize CMS}~\cite{Chatrchyan2012} experiments have been published
for a range of charged hadrons. Compared with $R_{AA}$ of hadrons (charged particles, $\pi,k,p $, etc.), $R_{AA}$ of light (anti)nuclei is
not well explained in high energy A-A collision experiments. Therefore we think the properties of $R_{AA}$ of (anti)hadrons and (anti)nuclei in Pb-Pb
collisions deserve to be further discussed in models.

Presently, there are many successful phenomenological models widely
used to describe the production of hadrons and light nuclei in relativistic
heavy-ion collisions~\cite{PR2018,NPA987}, such as the Ultra-relativistic
Quantum Molecular Dynamics ({\footnotesize UrQMD}) approach~\cite{PPNP1998},
a Multiphase Transport ({\footnotesize AMPT}) model~\cite{PhysRevC.72.064901},
and the Simulating Many Accelerated Strongly interacting Hadrons ({\footnotesize SMASH}) approach~\cite{PhysRevC.94.054905}.
For the light (anti)nuclei production in terms of their yields, yield ratios, spectra, flow, etc, coalescence models and statistical thermal method are usually employed,
as has been done in the frameworks of either the coalescence approach~\cite{Zhou2016,SHAH20166,nst28p,PhysRevC.92.064911,prc98w,plb792k,cpc43r,plb805h}
or the statistical model approach~\cite{plb785v,natu561a,prc99d,prc995f}.

In this paper, the production and transverse momentum ($p_{T}$) of final state
(anti)hadrons ($\pi^{+},\pi^{-}$, $p,\bar{p}$) are simulated by the {\footnotesize PACIAE} model~\cite{sa2012paciae} in $pp$ and Pb-Pb
collisions at $\sqrt{s_{\rm{NN}}}$ = 2.76 TeV. And then the dynamically constrained phase-space coalescence ({\footnotesize DCPC}) model~\cite{PhysRevC.85.024907} is
applied to deal with the production and properties of light (anti)nuclei ($d,\bar{d}$). Previous results of light (anti)nuclei
production for both $pp$~\cite{PhysRevC.85.024907,JiangThe} and A-A~\cite{PhysRevC.86.054910,PhysRevC.88.034908,gang2014scaling,Zhilei2016,Dong2018,PhysRevC.99.034904,Liufx2019} collisions
in relativistic energy region, including transverse momentum distribution, energy dependence, scaling property,
centrality dependence have been obtained using this framework.
In the rest of this paper, we will investigate the properties of nuclear
modification factors ($R_{AA}$) of (anti)hadrons and (anti)deuteron in Pb-Pb collisions
at $\sqrt{s_{\rm{NN}}}$ = 2.76 TeV with the same approach.

The paper is organized as follows: In sect. II, we concisely introduce
the {\footnotesize PACIAE} and {\footnotesize DCPC } model. In sect. III, our
numerical calculation results of the $R_{AA}$ for (anti)hadrons
and (anti) deuteron are presented and compared with the available experimental
data at {\footnotesize LHC}. In sect. IV, a brief summary is provided.

\section{Models}
\label{sec1}

The {\footnotesize PACIAE} model~\cite{sa2012paciae} based on {\footnotesize PYTHIA }6.4~\cite{Sj_strand_2006},
is designed and expanded to be feasible for p-p, p-A and
A-A collisions. In this model, the entire collision process can be mainly decomposed into four stages as follows:

Firstly, the partonic initial states are created. The nucleus -nucleus collision
can be simplified into numerous nucleon-nucleon ($NN$) collisions
according to the collision geometry and $NN$ total cross section.
Each $NN$ collision is described by the {\footnotesize PYTHIA} model
generating quarks and gluons for further evolution.
A partonic initial state of a nucleus-nucleus collision can be created when
all $NN$ collisions are exhausted. This state is also considered
as the quark-gluon matter ({\footnotesize QGM}) generated
in high energy nucleus-nucleus collisions. Secondly, the parton rescattering proceeds via the 2$\rightarrow$ 2 parton-parton scattering described by
the LO- pQCD cross sections~\cite{COMBRIDGE1977234}. Here,
a $K$ factor is added to include non-perturbative {\footnotesize QCD}
and higher-order corrections.
Thirdly, the hadronization process is treated through the Lund string fragmentation approach~\cite{Sj_strand_2006} or
the phenomenological coalescence method~\cite{sa2012paciae}.
Finally, the hadron rescattering is carried out till the exhaustion of hadron-hadron collision pairs or the hadronic freeze-out. One refers to~\cite{sa2012paciae} for the detail.

Then the production of light (anti)nuclei can be calculated with
the {\footnotesize DCPC} model~\cite{PhysRevC.85.024907} when the final state hadrons have already been provided by {\footnotesize PACIAE}.
Due to the uncertainty principle $(\Delta\vec q\Delta\vec p \geq h^{3})$,
one cannot simultaneously obtain the precise information of both position
$\vec{q\equiv} (x,y,z)$ and momentum $\vec{p\equiv} (p_x,p_y,p_z)$ for a particle
in the six-dimension phase space. Thus one can only deduce that this particle
lies in a quantum "box" with a phase-space volume of $\Delta\vec q\Delta\vec p$. Hence
we can simulate the yield of a single particle using an integral:
\begin{equation}
Y_1=\int_{H\leq E} \frac{d\vec qd\vec p}{h^3},
\end{equation}
where $H$ and $E$ denote the Hamiltonian and energy of the particle, respectively.
Analogously, one can compute the yield of the synthetic (anti)nuclei
containing N particles with the following integral:
\begin{equation}
Y_N=\int ...\int_{H\leq E} \frac{d\vec q_1d\vec p_1...d\vec
q_Nd\vec p_N}{h^{3N}}. \label{funct1}
\end{equation}
Note that, two constraint conditions have to be satisfied in this equation:
\begin{equation}
m_0\leq m_{inv}\leq m_0+\Delta m,
\end{equation}
\begin{equation}
|q_{ij}|\leq D_0,(i\neq j;i,j=1,2,\ldots,N)
\end{equation}
where
\begin{equation}
m_{inv}=\Bigg[\bigg(\sum^{N}_{i=1} E_i \bigg)^2-\bigg(\sum^{N}_{i=1}
\vec p_i \bigg)^2 \Bigg]^{1/2},
\end{equation}
and $E_i$, $\vec p_i (i=1,2,\ldots,N$) represent the energy and momentum of
one particle, respectively. $m_0$ and $\Delta m$ denote the rest mass of
synthetic (anti)nuclei and the allowed mass uncertainty. $D_0$ refers to
diameter of (anti)nuclei, and $|q_{ij}|$ stands for the vector distance
from $i$-th and $j$-th particles. The integration in Eq.~(\ref{funct1})
should be replaced by the summation over discrete distributions,
as a coarse graining process in the transport model.

\section{Results and Discussions}
\label{sec2}

\begin{figure}[t]
\includegraphics[width=0.43\textwidth]{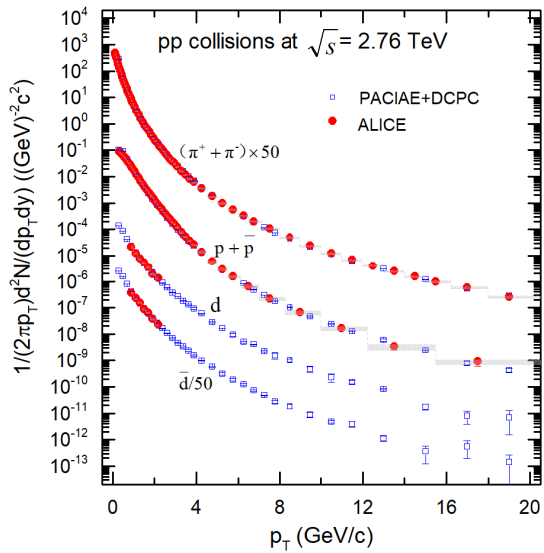}
\caption{(Color online) The transverse momentum spectra of charged pions, (anti)protons, and (anti)deuteron computed by {\footnotesize PACIAE+DCPC}
model (the open symbols) in $pp$ collisions at $\sqrt{s}= 2.76$ TeV, compared with {\footnotesize ALICE} results~\cite{PhysRevC.93.034913, PRC.97024615} (the solid symbols). The vertical lines (error bars) show the statistical uncertainty and the shaded areas represent the systematic
uncertainty of the {\footnotesize ALICE} results. The spectra have been scaled by the factors listed in the legend for clarity.}
\label{fig:1}
\end{figure}

\begin{figure*}[tb]
\centering
\includegraphics[width=0.86\textwidth]{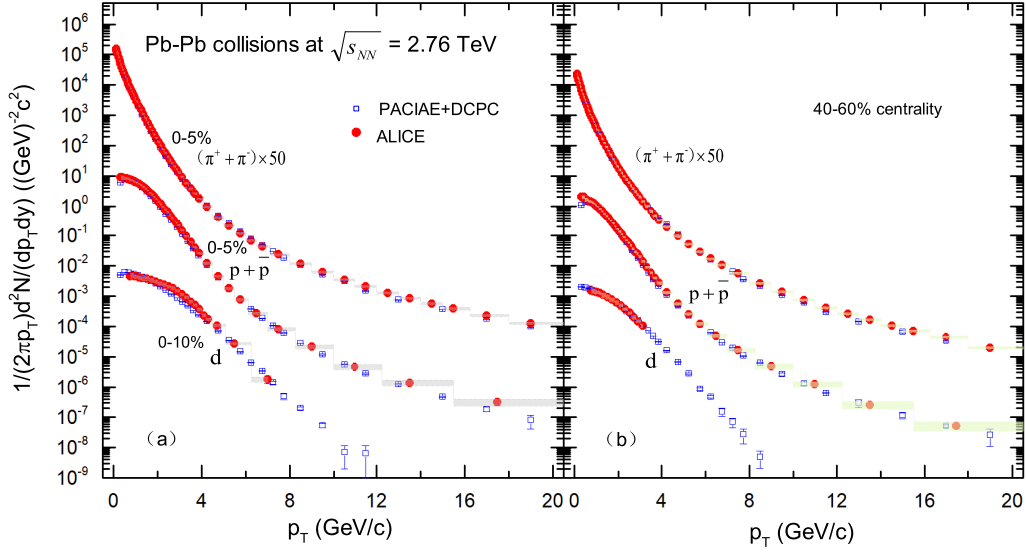}
\caption{The transverse momentum spectra of charged pions, (anti)protons, and deuteron
are presented by {\footnotesize PACIAE+DCPC} model (the open symbols) in Pb-Pb collisions at $\sqrt{s_{\rm{NN}}} = 2.76$~TeV,
compared with {\footnotesize ALICE} results~\cite{PhysRevC.93.034913, PRC93024917,EPJC2017658} (the solid symbols), (a) in
centrality bin of 0-5\% for $\pi^++\pi^-$, $p+\bar p$ and 0-10\% for $d$, (b) in centrality bin of 40-60\%, respectively. The
vertical lines (error bars) show the statistical uncertainty and the shaded areas represent the systematic uncertainty of
the experimental results. The spectra of charged pions have been scaled by the factors 50 for clarity.}
\label{fig:2}
\end{figure*}

At first, we can obtain the final-state particles in $pp$ and Pb-Pb collisions
using the {\footnotesize PACIAE} model~\cite{sa2012paciae}. In this simulation,
the hadrons are created on the assumption that hyperons heavier than $\Lambda$ are
already decayed, and most of model parameters are fixed on the default values
given in {\footnotesize PYTHIA}6.4~\cite{Sj_strand_2006}.
We determine the $K$ factor, parj(1,2,3) for primary hadrons in {\footnotesize PACIAE} model
by fitting to the {\footnotesize ALICE} pions and protons $p_T$ spectra data~\cite{PhysRevC.93.034913}.
The fitted values of $K$ = 2, parj(1) = 0.15, parj(2) = 0.50, and parj(3) = 0.60 for $pp$ collisions as well as
$K$ = 2, parj(1) = 0.15, parj(2) = 0.38, and parj(3) = 0.65 for Pb-Pb collisions are used in later calculations.
Here, parj(1) is the suppression of diquark-antidiquark pair production compared with the quark-antiquark pair production,
parj(2) is the suppression of strange quark pair production compared with $u$ ($d$) quark pair production,
and parj(3) is the extra suppression of strange diquark production compared with the normal suppression of a strange quark.
Then we generate charged pions and (anti)protons transverse momentum spectra by {\footnotesize PACIAE} model
with $|y|< 0.5$ and $0 < p_T < 20$~GeV/c at $\sqrt{s_{\rm{NN}}}$ = 2.76 TeV, in $pp$ collisions
as shown in Fig.~\ref{fig:1} and Pb-Pb collisions for centrality bin of 0-5\% and 40-60\%
as shown in Fig.~\ref{fig:2}, respectively.

Then the yields and transverse momentum spectra of (anti)deuteron were calculated by the dynamically
constrained phase-space coalescence model ({\footnotesize DCPC}) in $pp$ and Pb-Pb collisions at $\sqrt{s_{\rm{NN}}}$ = 2.76 TeV
according to the final hadronic states from the {\footnotesize PACIAE} model. Here, we choose the model parameter $D_0= 3$ fm
and $\Delta m$ = 0.42 MeV/c in $pp$ and Pb-Pb collisions~\cite{Zhilei2016}. In the end, we can compare the model calculations of
the nuclear modification factors for (anti)hadrons and light (anti)nuclei in Pb-Pb collisions at $\sqrt{s_{\rm{NN}}}$ = 2.76 TeV to
experimental data and study the quenching effect in relativistic heavy ion collisions.

In Fig.~\ref{fig:1}, the transverse momentum spectra of charged pions, and (anti)protons computed by {\footnotesize PACIAE}
model (the open symbols) in $pp$ collisions at $\sqrt{s}$ = 2.76 TeV within rapidity $|y| < 0.5$ were used to fit model parameters
with {\footnotesize ALICE} results~\cite{PhysRevC.93.034913} (the solid symbols). In addition, the transverse momentum spectra
of (anti)deuteron calculated by the {\footnotesize PACIAE+DCPC} model simulation (the open symbols) in $pp$ collisions at $\sqrt{s}$ = 2.76 TeV within
rapidity $|y| < 0.5$ are also shown in the Fig.~\ref{fig:1}, which is in good agreement with the known {\footnotesize ALICE} results~\cite{PRC.97024615}.

Similarly, Fig.~\ref{fig:2} shows the transverse momentum spectra of charged pions, and (anti)protons calculated
by {\footnotesize PACIAE+DCPC} model (open symbols) in Pb-Pb collisions at $\sqrt{s_{\rm{NN}}} = 2.76$ TeV for different
centrality bins of 0-5\% and 40-60\% within rapidity $|y| < 0.5$ confronted with {\footnotesize ALICE}
results~\cite{PhysRevC.93.034913} (the solid symbols). One can see from Fig.~\ref{fig:2} that for $p_T < 3.0$ GeV/c, the spectra
in central collisions becomes harder and there is a mass dependent effect. Both protons and pions $p_T$ spectra are well described
by our model in different centrality bins.
Then the transverse momentum spectra of deuteron computed by the {\footnotesize PACIAE+DCPC} model
simulation (the open symbols) in Pb-Pb collisions at $\sqrt{s_{\rm{NN}}} = 2.76$ TeV in both central and peripheral collisions
are in good agreement with the {\footnotesize ALICE} data~\cite{PRC93024917,EPJC2017658} as shown in Fig.~\ref{fig:2}.

\begin{figure*}[!htb]
\centering
\includegraphics[width=0.86\textwidth]{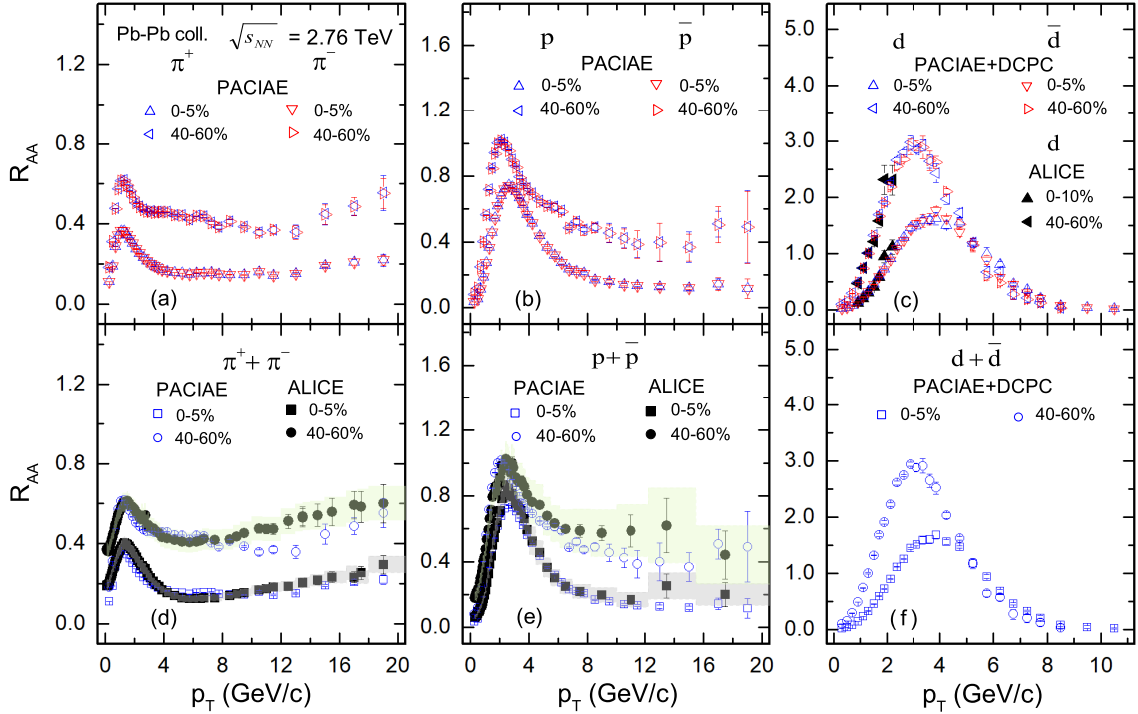}
\caption{(Color online)The nuclear modification factor $R_{AA}$ are calculated by {\footnotesize PACIAE+DCPC} model (the open symbols)
for different particle species in 0-5\% most central and 40-60\% peripheral Pb-Pb collision events at $\sqrt {s_{NN}}=2.76$~TeV,
as a function of $p_{T}$. The {\footnotesize ALICE} results (the solid markers) for comparison were taken from  Ref.~\cite{PhysRevC.93.034913}
for panel (d) and (e), and were computed using the data from Ref.~\cite{PRC.97024615,PRC93024917} for panel (c). The vertical lines (error bars)
show the statistical uncertainty and the shaded areas represent the systematic uncertainty of the experimental results.}
\label{fig:3}
\end{figure*}

\begin{figure*}[tb]
\centering
\includegraphics[width=0.75\textwidth]{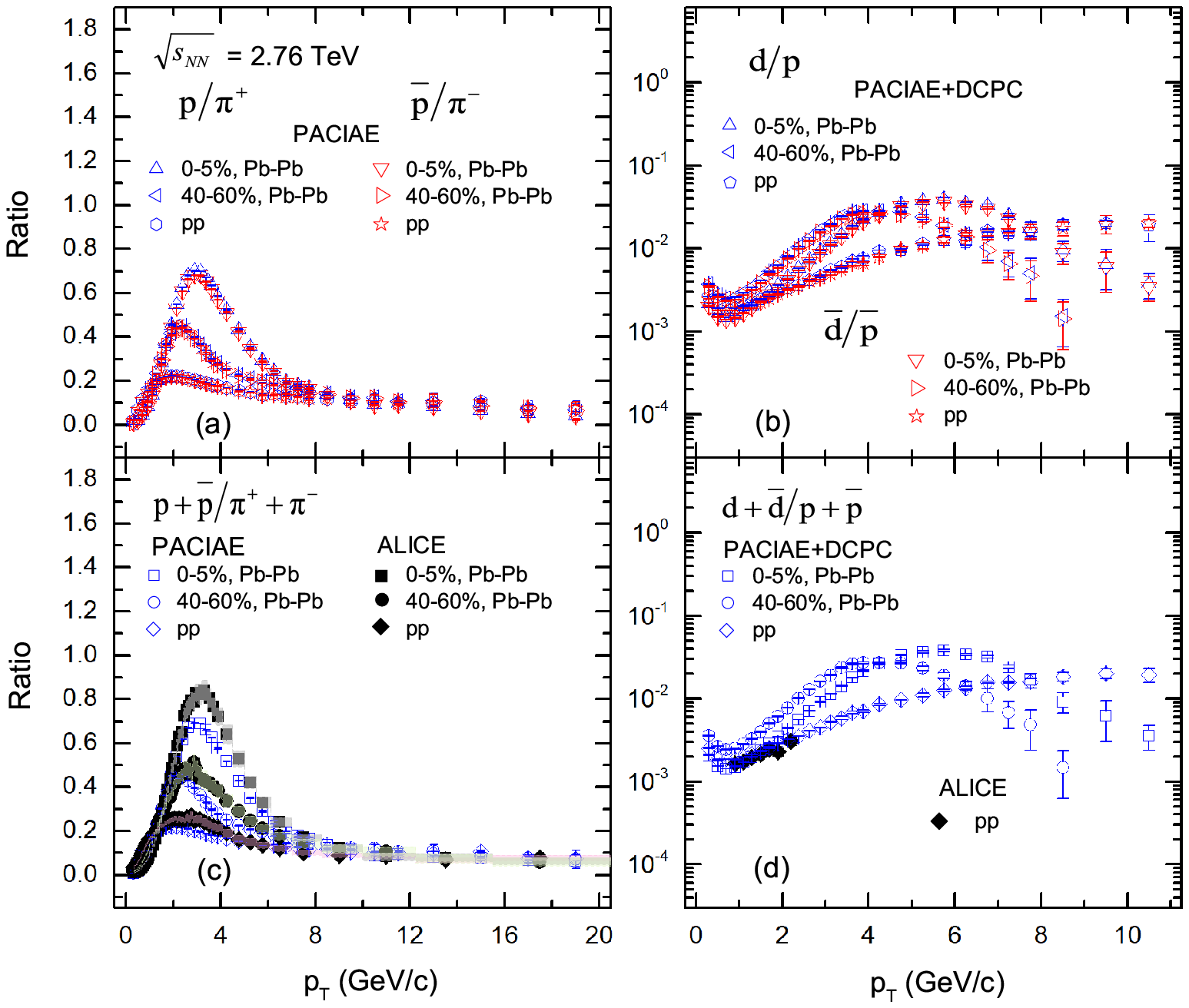}
\caption{The ratios of (anti)proton to charged-pion and (anti)deuteron to (anti)proton computed by {\footnotesize PACIAE+DCPC}
model (the open symbols) as a function of $p_{T}$ in $pp$ collisions, as well as the most central (0-5\%) and peripheral (40-60\%) Pb-Pb
collisions at $\sqrt {s_{NN}}=2.76$~TeV, respectively.
Here, {\footnotesize ALICE} results (the solid markers) for comparison were take from Ref.~\cite{PhysRevC.93.034913} in panel (c), and
were computed with the data from Ref.~\cite{PhysRevC.93.034913,PRC.97024615} in panel (d). The vertical lines (error bars) show
the statistical uncertainty and the shaded areas represent the systematic uncertainty of the experimental results.}
\label{fig:4}
\end{figure*}

The nuclear modification factor $R_{AA}$ for pion, proton and deuteron is shown in Fig.~\ref{fig:3} (the open symbols).
Figure~\ref{fig:3} (a) to (c) show the distribution of the nuclear modification factor $R_{AA}$ for the $\pi^+$, $p$, and $d$ compared to
their antiparticles $\pi^-$, $\bar{p}$, and $\bar{d}$, in two different centrality bins. Figure~\ref{fig:3} (d) to (f) show
the distribution of $R_{AA}$ versus $p_T$ for combined $\pi^+ +\pi^-$, $p+\bar p$, and $d+\bar d$.

From Fig.~\ref{fig:3}, one can see that the distribution
of the nuclear modification factor $R_{AA}$ for different particle species and different centrality increases
with $p_T$ value, reaches a peak, and then decreases with transverse momentum $p_T$, indicating a unified energy loss mechanism
is acting on all the different particle species including nuclei at high transverse momentum. And the depression effect of
central collision event are more significant than that of peripheral collision, due to a stronger medium modification effect
in central collisions. Next, we can see from Fig.~\ref{fig:3} (a) to (c) that the $R_{AA}$ distribution of
antihadrons and antinuclei are the same with that of corresponding hadrons and nuclei, showing that
the $R_{AA}$ suppression or quenching effect on matter and antimatter is the same in high energy Pb-Pb collisions.
It is worth noting, as shown in Fig.~\ref{fig:3} (c) and (f), that the suppression or quenching effect
in the high transverse momentum region is more significant for nuclei than in meson and baryons.

The solid markers in Fig .~\ref{fig:3} (c), (d), and (e) represent the experimental data~\cite{PhysRevC.93.034913,PRC.97024615,PRC93024917} compared
with our simulation results.
It is observed that the $R_{AA}$ results of the $\pi^+ +\pi^-$, $p+\bar p$ and $d$ from our simulation
are comparable to those of the {\footnotesize ALICE} data at $p_{T} <$ 10.0~GeV/c within the current errors
in Fig.~\ref{fig:3} (c), (d), (e); while as $p_{T} > 10.0$~GeV/c, our simulation is off the data by a small factor.
It should be mentioned that the {\footnotesize ALICE} data $R_{AA}(d)$ used for comparison in Fig.~\ref{fig:3} (c) were
calculated according to Eq.~(\ref{funct0}) based on the experimental data taken from Ref.~\cite{PRC.97024615} for $pp$ collisions
and Ref.~\cite{PRC93024917} for Pb-Pb collisions.

\begin{figure*}[htbp]
\centering
\includegraphics[width=0.76\textwidth]{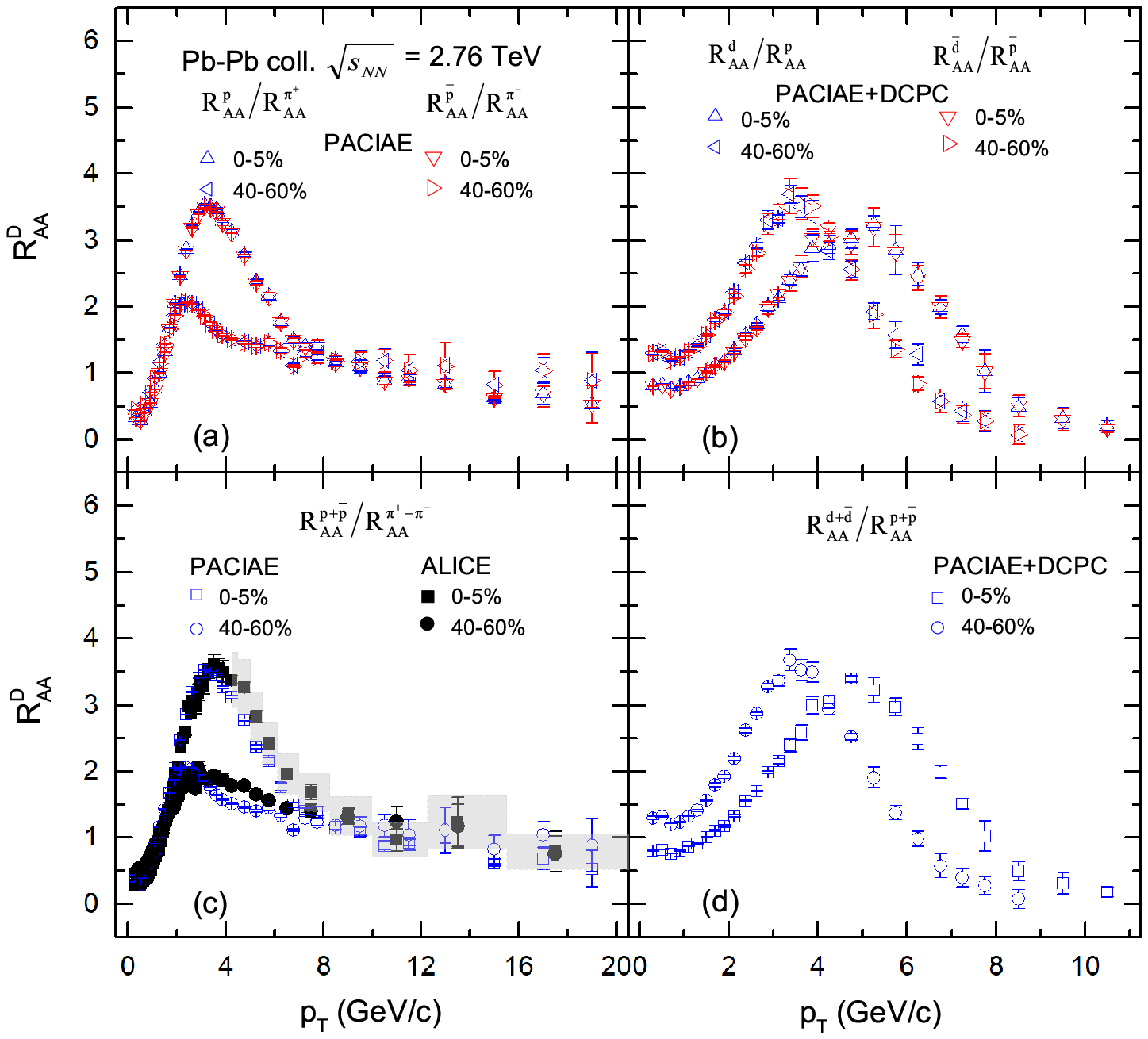}
\caption{ The double ratios $R^D_{AA}$ of (anti)proton to charged-pion and (anti)deuteron to
(anti)proton computed by {\footnotesize PACIAE+DCPC} model (the open symbols) as a function
of $p_{T}$ in $pp$ collisions, as well as in Pb-Pb collisions of the centrality bins
of 0-5\% and 40-60\% at $\sqrt {s_{NN}}=2.76$~TeV, respectively. Here, {\footnotesize ALICE}
data (the solid markers) for comparison in panel (c), at $p_{T}> $ 4.0~GeV/c, were taken
directly from Ref.~\cite{PLB2014196}; at $p_{T}< $ 4.0~GeV/c, were calculated using the data
from Ref~\cite{PhysRevC.93.034913}. The vertical lines (error bars) show the statistical uncertainty
and the shaded areas represent the systematic uncertainty of the experimental results.}
\label{fig:5}
\end{figure*}

We also perform a particle ratio study versus $p_T$ for (anti)proton to charged pion and (anti)deuteron to (anti)proton in this model.
Figure~\ref{fig:4} (a) and (b), display the ratio distributions of $p/\pi^{+}$, $\bar{p}/\pi^{-}$,
$d/p$, and $\bar{d}/\bar {p}$, respectively. It's easy to see that the distributions of the ratio
for $p/\pi^{+}, d/p$ are similar to $\bar{d}/\bar{p},\bar {p}/\pi^{-}$ in $pp$ collisions, central and
peripheral Pb-Pb collisions, suggesting a common suppression behavior for the matter and antimatter.

The ratio distributions of $p+\bar{p}/\pi^{+}+\pi^{-}$ and $d+\bar{d}/p+\bar{p}$ are shown
in Fig.~\ref{fig:4} (c) and (d). It can be seen that for the central and peripheral Pb-Pb collisions,
the ratio grows to a maximum value at $p_T \sim 3.0$ GeV/c for ($p+\bar{p}/\pi^{+}+\pi^{-}$)
and $p_T \sim 5.0$ GeV/c for ($d+\bar{d}/p+\bar{p}$), then decreases as $p_T$ increases.
In Fig .~\ref{fig:4} (c) and (d), the solid markers show the {\footnotesize ALICE} results~\cite{PhysRevC.93.034913} for comparison.
Obviously, the $p+\bar{p}/\pi^{+}+\pi^{-}$ ratio in our simulation shows a similar structure to that in data.
The ALICE data ($d+\bar{d}/p+\bar{p}$) used for comparison in Fig.~\ref{fig:4} (d) were computed using data ($p+\bar{p}$) taken
from Ref.~\cite{PhysRevC.93.034913} and data ($d+\bar{d}$) from Ref.~\cite{PRC.97024615}.

To quantify the similarity of the suppression, the double $R^D_{AA}$ ratio were defined, such as
the double ratio $R^D_{AA}$ of protons to pions is defined as follows~\cite{PLB2014196}:

\begin{equation}
R^D_{AA} = \frac{R^{p+\bar{p}}_{AA}}{R^{\pi^{+}+\pi^{-}}_{AA}},
\label{funct7}
\end{equation}
where $R^{\pi^{+}+\pi^{-}}_{AA}$ and $R^{p+\bar{p}}_{AA}$ denote the $R_{AA}$ for
the charged pion and proton, respectively. This double ratios constructed using the particle ratios may be
properly handled that the dominant correlated systematic uncertainties are
between particle species and not between different collision systems.

Fig.~\ref{fig:5} shows the double $R^D_{AA}$ ratios of
protons ($p, \bar p, p+\bar p$) to pions ($\pi^+ ,\pi^-, \pi^+ +\pi^-$) and deuterons ($d, \bar d, d+\bar d$) to
protons ($p, \bar p, p+\bar p$), as a function of $p_T$, calculated
by {\footnotesize PACIAE+DCPC} in the most central (0-5\%)
and peripheral (40-60\%) Pb-Pb collisions at $\sqrt {s_{NN}}=2.76$~TeV, respectively.
We can see from Fig.~\ref{fig:5}, that the $R^D_{AA}$ for all particle combinations are generally
increasing at low $p_T$ and decreasing at high $p_T$. And comparing Fig.~\ref{fig:5} (a),(c) with
Fig.~\ref{fig:5} (b),(d) , we can also conclude that the suppression effect of the double $R^D_{AA}$ ratio of
deuteron to proton is more significant than that of proton to pion, as $p_T > 8$~GeV/c. Besides,
it is clear that, as shown in Fig.~\ref{fig:5} (a) and (b), the distribution of the double $R^D_{AA}$ ratios
for $p$ to $\pi^{+}$ and $d$ to $p$ are the same as that of
corresponding antimatter $\bar {p}$ to $\pi^{-}$ and $\bar {d}$ to $\bar {p}$ , which indicates that
matter and corresponding antimatter have the same suppression characteristics. Meanwhile,
from Fig.~\ref{fig:5} (c) it can be seen that the distribution of the results $R^D_{AA}$ from computed
by model simulation are consistent with the {\footnotesize ALICE} data~\cite{PLB2014196,PhysRevC.93.034913}.
It should be noted that the experimental values of double ratios $R^{p +\bar {p}}_{AA}/R^{\pi^{+} + \pi^{-}}_{AA}$ used
for comparison in Fig.~\ref{fig:5} (c), when $p_{T}< $ 4.0~GeV/c, were calculated
using data $R_{AA}^{p+\bar {p}}$ and $R_{AA}^{\pi^{+} + \pi^{-}}$ taken from
Ref.~\cite{PhysRevC.93.034913}, and when $p_{T}> $ 4.0~GeV/c, were taken directly from Ref.~\cite{PLB2014196}.

\section{Conclusions}
\label{sec3}

In the paper, we have studied the transverse momentum (${p_{T}}$) spectra of charged
particles $\pi^+ +\pi^-$ and $p+\bar p$ at scaled midrapidity $|y| < 0.5$ in $pp$ collisions,
in most central (0-5\%), and peripheral (40-60\%) Pb-Pb collisions by {\footnotesize PACIAE} model.
The key model parameters are determined by fitting pion and proton $p_T$ spectra data. The ${p_{T}}$ spectra
of deuteron ($d, \bar d$) are also simulated in this work using the {\footnotesize PACIAE + DCPC} model.
Then, the nuclear modification factors ($R_{AA}$) of charged pions, (anti)protons, and (anti)deuteron,
as well as, their yield ratios, double $R^D_{AA}$ ratios with $|y|<0.5$ in peripheral (40-60\%) and
central (0-5\%) Pb-Pb collisions at $\sqrt {s_{NN}}=2.76$~TeV have been studied using
the {\footnotesize PACIAE + DCPC} model. It is found that the $R_{AA}$ distribution of
light (anti)nuclei ($d, \bar d$) is similar to that of hadrons ($\pi^\pm, p, \bar p$), and
the $R_{AA}$ of anti-particles is the same as that of particles.
The suppression of $R_{AA}$ at high-$p_T$ strongly depends on event centrality and mass of
the particles.

It is interesting that there are no differences in nuclear modification between particles
and antiparticles in this work. In the {\footnotesize PACIAE} and {\footnotesize DCPC} models,
there is no equilibrium assumption between particles and antiparticles. It simulates
dynamically the whole relativistic heavy-ion collision process from the initial partonic
stage to the hadronic final state via the parton evolution, hadronization, and
hadron evolution according to copious dynamical ingredients (assumptions) introduced reasonably.
Therefore it is parallel to the experimental nucleus-nucleus collision. These dynamics correctly
describe the particle, energy, and entropy. Messages brought by the produced particles
in these transport (cascade) models are all dynamically generated. We do not apply
any equilibrium condition in our study and therefore sees no particle and antiparticle
difference in the simulation. Of course, further studies are required
to model the system evolutions in more sophisticated ways.

Most of the results predicted by our theory model are consistent with existing experimental results,
while others are somewhat different, such as the $R_{AA}$ distribution of charged pions
at the high-$p_T$. Therefore, it is necessary to improve the model.

\section {ACKNOWLEDGMENT}
This work was supported by the NSFC(11475149, 11775094, 11905188), as well as support
from the high-performance computing platform of China University of Geosciences.
The authors thank Prof. Che-Ming Ko for very helpful discussions.

\end{document}